# Dynamic, viscoelasticity-driven shape change of elastomer bilayers

Wenya Shu, C. Nadir Kaplan[*], and Justin R. Barone[*]

Biological Systems Engineering
Physics
Center for Soft Matter and Biological Physics
Macromolecules Innovation Institute
Virginia Tech
Blacksburg, VA 24018, USA
E-mail: nadirkaplan@vt.edu
E-mail: jbarone@vt.edu

**Abstract**

Thin bilayers made of elastic sheets with different strain recoveries can be used for dynamic shape morphing through ambient stimuli, such as temperature, mass diffusion, and light. As a fundamentally different approach to designing temporal shape change, constituent polymer molecular features (rather than external fields) are leveraged, specifically the viscoelasticity of gelatin bilayers, to achieve dynamic three-dimensional (3D) curls and helical twists. After stretching and releasing, the acquired 3D shape recovers its original flat shape on a timescale originating from the polymer viscoelasticity. The bilayer time-dependent curvature can be accurately predicted from hyperelastic and viscoelastic functions using finite element analysis (FEA). FEA reveals the nonlinear shape dynamics in space and time in quantitative agreement with experiments. The findings present a new frontier in dynamic biomimetic shape-morphing by exploiting intrinsic material properties in contrast with state-of-the-art methods relying on external field variations, moving one step closer to acquiring autonomous shape-shifting capabilities of biological systems.

# 1. Introduction

Natural heterogeneous structures display shape change as part of biological processes. Pine cone scales have two layers of different water swelling ability that allow them to remain closed in a wet state on the tree and to open and disperse seeds when they are off the tree and dry out.[1] Seed pods have two halves with cellulose fibers oriented 45º to the seed pod long axis and orthogonal to each other, which remain closed while wet on the plant.[2] Once off the plant, the halves dry out, and the fiber anisotropy causes twisting that opens the pod and releases the seeds. Another example is a cucumber tendril, which grows around objects in a helix towards light.[3] The twist arises from the heterogeneity originating in half of the tendril's cross-section being lignified. In each case, time-dependent shape change arises from mass diffusion into and/or out of the structure.

Structural heterogeneity is a biological design principle ripe for biomimicry to create synthetic devices and "origami"-like materials. Two polymer layers of different swelling ability formed into a bilayer makes the simplest shape-shifting device that mimics the pine cone scale. The polymer layers are typically hydrogels where the active layer swells more than the passive layer resulting in bending towards the passive layer. The swelling can be solvent (i.e., water)-, temperature-, pH-, electric-field- or ionic strength-induced.[4] Folding into more sophisticated final shapes can be achieved via photopatterning, lithography, or 3D printing (for 4D printing).[5] Seed-pod-like twisting has also been mimicked by imparting orientational order into polymer and elastomeric bilayers to generate solvent- or temperature-induced twisting, as well as pH-, ionic strength-, and UV light-induced shape change.[6] An interesting variant is a system with cross-sectional heterogeneity like the cucumber tendril and the ability to preferentially align fibers to affect helical pitch like the seed pod.[7] The main drawback of these stimulus-sensitive designs is that faster temporal responses require thinner bilayers due to the diffusion limit of heat or solvent as well as the penetration depth of light.

One way to overcome this limitation is to use a strain-induced stimulus or "mechanical programming", which has been applied to bilayers with fiber orientation variation between the two layers similar to the seed pod.[8] Systems that respond to an applied deformation do not necessarily need to be of a limited size. The cucumber tendril also shows that a responsive shape-changing bilayer does not need orientational anisotropy. Thus, bilayers with heterogeneous elastic and plastic properties through the cross-section can be constructed that can undergo out-of-plane deformations in response to mechanical stimuli and then arrest the final shape. Simple designs have used elastic rubber bilayers for permanent shape change: one layer of rubber is stretched and then bonded to an unstretched, more rigid layer. Upon release, the unstretched layer restricts the relaxation of the stretched layer, leading to permanent bending, curling, and twisting with a pitch variation.[9] Other systems, such as polyolefin elastomer or rubber-plastic bilayers have exploited the ratio of the elastic to plastic response to form stable helices, tubes, and saddles without the need of any orientation imprinting.[10]

Mechanically-induced shape change must ideally be augmented to acquire a tailored temporal response for reversibly actuating systems in analogy with their biological counterparts. However, whereas the light fields and diffusion-controlled stimuli strongly limit the material dimensions to achieve reasonable deformation rates[11], the strain-induced stimuli in elastomeric bilayers lead to impractically fast shape changes, such as a rubber band rapidly snapping back upon pulling and releasing. The fundamental material property to circumvent these restrictions is viscoelasticity, which enables a time-dependent mechanical response. Dynamic shape changes have been realized through the selective spatial relaxation of stretched viscoelastic polystyrene (PS) sheets above the glass transition.[12] Hydrogel bilayers with varying viscoelasticity can achieve time-dependent shape change that can be tailored with solvent or light stimulus.[13] To that end, we hypothesize that viscoelastic layers with different time-dependent properties could form shape-morphing materials that bend, curl, and twist on timescales determined solely by the intrinsic polymer molecular structure.

To validate this hypothesis, we synthesize flat quasi-1D bilayers from gelatin layers with different viscoelasticity that can attain 3D shapes when stretched and released. Over time, the 3D shapes return to their original flat conformation. The viscoelastic recovery process is associated with the intrinsic polymer molecular diffusion that can be controlled through polymer molecular design. We use a simple theory based on the Timoshenko bilayer model as well as more comprehensive FEA simulations to explain the geometrical dynamics of the system. The experiments and simulations together pave the way for the synthesis of shape-shifting materials designed completely at the molecular scale by using layers without inherent anisotropy at rest that can achieve very high curvatures.

## 2. Results and Discussion

The bilayers are stretched to $\varepsilon_a$=0.25, 0.50, 0.75, and 1.00 cm cm$^{-1}$ applied strain and released (a time lapse sequence is shown in Figure 1). Immediately upon release, the bilayer attains its highest curvature at zero time, $\kappa_0$. Initial curvature, $\kappa_0$, increases with increasing applied strain (Figure 2).[10a] Low curvature bending, i.e., $\kappa_0 < 2\pi/L$, is observed at both water contents for $\varepsilon_a$=0.25 cm cm$^{-1}$. Higher curvature curling, i.e., $\kappa_0 \geq 2\pi/L$, is observed for the lower water content at $\varepsilon_a$=0.50 cm cm$^{-1}$ and the higher water content at $\varepsilon_a$=0.75 cm cm$^{-1}$ (Figure 3a). The higher water content at $\varepsilon_a$=0.50 cm cm$^{-1}$ shows a mixture of bending and curling and has the highest standard deviation in the curvature. Finally, both water contents at $\varepsilon_a$=1.00 cm cm$^{-1}$ show twisting into helices, i.e., $\kappa_0 >> 2\pi/L$. The lower water content at $\varepsilon_a=0.75$ cm cm$^{-1}$ mostly shows helices but occasionally curls and has a higher standard deviation. For swollen bilayers, bending happens because the active layer expands or swells more than the passive layer resulting in bending towards the thinner passive layer. The swollen layer is in tension while the unswollen layer is in compression. For differently-strained bilayers, the shorter layer (ER-high) is in tension because it is being pulled by the longer layer (ER-low), which is in compression, resulting in bending towards the shorter ER-high layer.[10a, 14]

To describe the stretching and releasing of a quasi-1D bilayer to achieve a 3D shape, a modified creep recovery experiment is used to determine the strain difference between the two layers at any point in time (Figure 4). After stretching and releasing, each layer has an immediate and equal elastic recoil and then there is an initial strain difference, $\Delta\varepsilon_0$, between the ER-low, $\varepsilon_l$, and ER-high, $\varepsilon_h$, layers in the bilayer, $\Delta\varepsilon_0=\varepsilon_{l,0}-\varepsilon_{h,0}$, where the "0" subscript denotes the zero-time behavior or the point at which a strain difference emerges between the two layers of the bilayer. This point is defined as the transition from step (3) to step (4) in Figure 4. At this point and beyond, the ER-high layer is always shorter than the ER-low layer, resulting in bending towards the ER-high side. The glycerol-plasticized ER-high layer exhibits high elastic recovery and minimal creep, allowing it to quickly regain its shape after deformation. In contrast, the glycerol-less ER-low layer demonstrates more viscous behavior and lower elastic recovery. The initial curvature, $\kappa_0$, is easier to understand when plotted as a function of the initial strain difference, $\Delta\varepsilon_0$, where the data collapses onto 1 curve (Figure 3b, linear fit of $r^2=0.89$ represented by the black line). Initial strain differences of $\Delta\varepsilon_0 \leq 0.15$ result in bending, $\Delta\varepsilon_0=0.20-0.30$ result in curling, and $\Delta\varepsilon_0 \geq 0.4$ result in twisting into helices.

After attaining the initial maximum value, the curvature decreases over time until the bilayer is nearly flat again. The experimentally observed curvature shows two types of decay: a fast decay for the bilayers with more water and a slow decay for the bilayers with less water (Figure 5). The bilayer curvature change with time is envisioned as a viscoelastic creep recovery process where the strain difference between the two layers with time, $\Delta\varepsilon(t)=\varepsilon_l(t)-\varepsilon_h(t)$, results in the curvature change with time, $\kappa(t)$. When the bilayer is stretched and released, the mismatch in residual strain between the layers creates high shear stress at the interface (Figure S4). Consequently, the bilayer bends toward the plasticized ER-high layer, which has less residual strain after release. Over time, the residual strain difference between the layers decreases due to polymer molecule relaxation, resulting in a gradual reduction of internal stresses within the bilayer caused by the initial strain mismatch. As a result, the bilayer experiences a progressive reduction in curvature, gradually returning toward its initial straight shape. Constructing the two layers out of the same material allows for a strong interface to transfer stress across with no debonding and is an important consideration in the design.

The bilayer curvature is a function of the strain difference between the layers and geometric/material parameters, i.e., $\kappa=f(\Delta\varepsilon)$. The Timoshenko bilayer model linearly relates the bending curvature of a bimetal strip to the mismatched thermal strains in its different layers and works well for a quasi-1D geometry. Here, we replace the thermal strain difference between the layers in the original Timoshenko formulation by the strain difference between the two layers obtained from the modified creep recovery experiments, $\Delta\varepsilon$, which yields:[14]

$$\kappa(t) = \frac{6[\Delta\varepsilon(t)](1+s)^2}{H\left[3(1+s)^2 + (1+sn)\left(s^2 + \frac{1}{sn}\right)\right]} \quad (1)$$

In the Timoshenko theory and its extension to solvent swollen bilayers[1a], s is the layer thickness, H, ratio and n is the layer modulus, µ, ratio. Bilayers always bend towards the ER-high layer because it has less residual strain (or more elastic recovery) than the ER-low layer, thus $s=H_h/H_l$ and $n=\mu_h/\mu_l$, where l and h denote the ER-low and ER-high layers, respectively, and $H=H_h+H_l$ is the total bilayer thickness. Since the bilayer is stretched and released, the actual bilayer thickness will change with the strain, ε, as $H=H_0/(\varepsilon+1)^{1/2}$, where $H_0$ is the original bilayer thickness, and the adjusted thickness is used in the s calculation. While the ratio of elastic modulus, n, may vary with time, we consider for this analysis a constant ratio of the low strain Hookean moduli with $n=\mu_0^h/\mu_0^l=0.25$, because its variation has limited influence on curvature.[1a, 14] The modulus is inversely identified by performing FEA of the uniaxial tensile test to reproduce the stress-strain data (Figure S2g,h). The Timoshenko bilayer model predicts the zero-time curvature, $\kappa_0$, at low $\Delta\varepsilon_0$ reasonably well but tends to overpredict it at higher $\Delta\varepsilon_0$ (Figure 3a,b).[1a, 14] This overestimation could be attributed to the assumption of small deformation and a state involving only modest bending in the Timoshenko bilayer. These assumptions limit the model accuracy in characterizing the gelatin bilayers, which experience both large deformation and torsion at high $\Delta\varepsilon$ (Figure S5 for the torsion of the bilayer). However, as the strain difference between the two layers and the torsion in the bilayers decreases, the time-dependent curvatures from the Timoshenko and FEA models, $\kappa(t)$, agree reasonably well (Figure 5).

Unlike the Timoshenko bilayer model, which considers only bending deformation and assumes small deformation conditions, nonlinear finite element analysis (FEA) offers a robust method for simulating both the curvature dynamics and the exact geometry of the bilayer within the context of large deformations. At large strains, the FEA simulations demonstrate improved accuracy in capturing the trend of zero-time curvature with strain differences compared to the Timoshenko model (Figure 3a,b). The FEA results also better capture the time-dependent curvatures of bilayers subject to significant initial stretches, particularly in the period after release ($\varepsilon_a \geq 0.5$ cm cm$^{-1}$, Figure 5). Furthermore, the FEA simulations accurately replicate the 3D shapes of the bilayer, including the highly twisted and helical deformations formed after high initial stretches (Figure 2c,d).

In the modified creep recovery experiment, the sample is stretched to a constant strain (that achieves a constant stress), held briefly, and released and the sample strain decay is monitored with time. After releasing the stress, the instrument crosshead must "chase" the recoiling sample, constantly moving down and intermittently pulling back to feel for a very small force to let the instrument know it is "chasing" at the proper rate. This is observed as some noise and a larger standard deviation in the low time regime. One advantage of the FEA prediction may lie in being able to use easier to obtain stress relaxation data to predict the strain recovery. In the stress relaxation experiment, the sample is stretched to a constant strain and held while the stress decay is monitored with time with much less error in the low time regime. From the hyperelastic stress-strain and viscoelastic stress relaxation data, the FEA can fairly accurately predict the strain recovery of each layer with time when subject to the same applied strain and hold procedure (Steps 1 and 2 in Figure 4) as the samples in the modified creep recovery experiment (Figure 6, plotted as the strain difference between the 2 layers). The FEA-predicted bilayer strain recovery difference, $\Delta\varepsilon(t)$, calculated from the Maxwell and Ogden models can be directly input into the Timoshenko bilayer model (Equation 1) to predict the curvature at modest applied strains (Section S4, Figure S6). This may be advantageous for initial bilayer design screening instead of performing strain recovery experiments on the individual layers and demanding nonlinear FEA on the bilayer itself.

The typical creep recovery experiment follows a creep test where the applied stress allows the sample to reach a steady-state strain rate before the stress is removed and the strain recovery monitored. The test is often performed well within the linear viscoelastic region, where the stress-relaxation, creep, and creep recovery can be related.[15] Since a linear viscoelastic stress relaxation model is used in the FEA (Equations S1, S2) it would be expected that in the low strain linear viscoelastic region the FEA could predict a typical creep recovery experiment. Here, the FEA model is able to adequately predict the modified creep recovery. The deviation in $\Delta\varepsilon_0$ and $\varepsilon(t)$ between the experiments and FEA gets a little larger at higher applied strains. Strain-dependent oscillatory shear experiments (at a frequency of $\omega=6.28$ rad s$^{-1}$) on these gelatin gels show the linear viscoelastic region to persist up to a strain of about 0.40 cm/cm, which is similar to other observations on gelatin gels of the same solids content.[16] A comparison of stress relaxation results using linear viscoelastic and non-linear viscoelastic relationships in the FEA shows that there is not much deviation up to applied strains of 1.00 cm cm$^{-1}$, indicating that the material responses are still within the (quasi)linear regime at the strains used in this study (Section S5). However, the deviations become larger at much higher strains albeit ones not used in this study.

The curvature is dependent on the strain difference between the two layers that persists with time (Figures 5, 6). When the ER-high layer is stretched to 100% strain and released, it recovers almost all the imposed strain immediately, about 93% in Figure 4 (or Figure S9d, where only the time-dependent strain recovery is shown), and then recovers the other 7% over time. This indicates that it is mostly elastic with a small viscoelastic component to its behavior. The stress relaxation response also shows this, where the ER-high layer behaves more like an elastic solid (Figure S2c,d). The stress decreases to about 70-80% of its initial value over time. On the other hand, the ER-low layer immediately recovers much less of the imposed strain (about 50% in Figure 4, S9d).

It then recovers the other 50% over time. To achieve the initial strain difference, $\Delta\varepsilon_0$, two layers with different elastic recoveries need to form the bilayer. Here, that is achieved by plasticizing the ER-high layer with glycerol, which has the physical effect of lowering its modulus and increasing its elastic recovery compared to the ER-low layer (Table S3).

The strain recovery difference is dominated by the viscoelasticity of the ER-low layer (Figure S9). Glycerol creates a gelatin elastomer with high elastic recovery (resilience) and low viscous loss in the ER-high layer. Glycerol solvates or "plasticizes" the majority of the gelatin molecules, reducing polymer inter-molecular friction and allowing fast strain recovery and more elastic behavior. The stress relaxation data shows that the ER-high layer actually relaxes slower than the ER-low layer (Figure S2c,d, Table S2). This arises because glycerol elicits slower polymer molecule relaxation modes compared to corn syrup.[16b] Corn syrup appears to promote more bonding between the gelatin and sugar molecules leading to a higher modulus. The faster relaxation in the ER-low layer seems to be shorter time Rouse modes, possibly from dangling polymer chain ends, that occur after a modest elastic recovery. The slower relaxation in the ER-high layer seems to be longer time reptation modes that occur after the large elastic recovery. The slower ER-high layer relaxation does not matter because the small amount of viscoelasticity exhibited by the ER-high layer is much less important than its large immediate strain or elastic recovery as the ER-low layer has to recover a lot more strain regardless of the rate. This shows that in designing layers for a bilayer, the plasticizer type can yield different viscoelasticity to fine tune the temporal response.

A bilayer with an intrinsic viscoelasticity difference between the two layers can have its viscoelasticity further changed by changing the overall plasticizer content of both layers. Here, that is achieved by changing the overall water content of the bilayers while keeping the glycerol and corn syrup contents constant. When the bilayers contain more water ($m/m_0$=0.90 or 0.114 mass fraction water bilayers) the curvature decay is faster than bilayers with less water ($m/m_0$=0.85 or 0.062 mass fraction water bilayers). In a viscoelastic system, the time scale for shape change is the polymer molecule diffusion time, $\tau$, which depends on the polymer molecule length, temperature, and inter-molecular friction coefficient. The polymer molecule diffusion time is $\tau \sim \ell^2/D$, where $\ell$ is a characteristic polymer molecule length (which is different for Rouse and reptation modes) and D is the polymer molecular self-diffusion constant. The diffusion constant assumes the Einstein form: $D \sim KT/\zeta$, where $\zeta$ is the polymer inter-molecular friction coefficient.[17] Different synthetic approaches can be used to change $\ell$ if desired to get a new $\tau$. All experiments are performed at room temperature but increasing temperature can make $\tau$ faster. Here, $\zeta$ is changed, $\tau \sim \ell^2 \zeta/KT$, so molecules with less water experience more friction and will diffuse slower. The 2 bilayers of different water content show the effect of changing the inter-molecular friction coefficient with higher water contents lowering $\zeta$ and showing faster strain and curvature decay. This appears in the viscoelastic response as lower $\tau$'s in the the Maxwell model (Table S2). The overall water content also has other effects. Observation of the strain recovery or curvature with time shows that they do not return to zero at long times as the applied strain increases or water content decreases. This seems to indicate some plastic deformation in the bilayer, largely from the ER-low layer.

Besides changing the polymer molecular features to alter the viscoelastic response, the rate of the applied strain to the bilayer can also be changed. FEA simulations show that a bilayer strained to the same extent (i.e., $\varepsilon_a$=1.00 cm cm$^{-1}$) at high strain rate will achieve a much lower curvature than a bilayer strained at low strain rate (Section S7, Figure S10). The polymer chains can respond to the applied strain and can stretch when strained slow enough but cannot when the strain is applied too fast.

The presence of small molecule liquid diluents or plasticizers means that there could be a poroelastic component to the time-dependence. For instance, water, glycerol, or dissolved sugar may diffuse through the crosslinked gelatin network. While that is true, the viscoelastic time scale is much shorter than any poroelastic time. The poroelastic time, $\tau_p$, dependends on the sample size as $\tau_p = H^2/D_s$, where H is the sample thickness and $D_s$ is the diffusion constant of the small molecule, i.e., water, glycerol or dissolved sugar, in the gelatin matrix. The diffusion constant of 10.4 wt% sucrose in water swollen gelatin at 25 °C is $D_s = 1.14 \times 10^{-6}$ cm$^2$ s$^{-1}$ (the value given in the reference is adjusted to 25 °C).[18] While this is not the exact condition of the gelatin here, i.e., the bilayers have much less water, it can be used in a conservative estimate of $\tau_p$ because there would be faster diffusion in more water. The diffusion constant of water in gelatin over the mass fraction range here (0.06-0.12 for $m/m_0$=0.85-0.90, respectively) is $D_s \sim 1 \times 10^{-8}$ cm$^2$ s$^{-1}$ at 25 °C.[19] This results in poroelastic relaxation times of $\tau_p = (0.42^2$ cm$^2)/D_s$=43-4900 *hours* (using the thinner sample size), which is consistent with what is observed in hydrogels.[20] Here, the small molecule components exist as plasticizers that hydrogen bond to the gelatin and alter the inter-molecular friction coefficient.[16b]

## 3. Conclusions

Time-dependent shape change can be realized using bilayers composed of two layers of different viscoelasticity. Flat bilayers are formed from two gelatin layers where one layer contains glycerol (ER-high) and the other does not (ER-low). After straining and releasing, the ER-high layer has much higher strain recovery than the ER-low layer. This creates a strain imbalance between the two layers that results in out of plane bending into a 3D shape. The elastomeric gelatin layers can achieve high strain differences resulting in high curvature helical twists ($\kappa \gg 2\pi/L$ where L is the bilayer length) that go well beyond simple bending. Over time, the 3D shape decays back to the original flat shape because of the intrinsic viscoelasticity of the ER-low layer that allows its strain to recover over time. Using simple polymer physics concepts, the viscoelasticity is changed by changing the polymer molecular environment, thus relating macroscopic shape change back to molecular features. Stress relaxation and modified creep recovery experiments phenomenologically describe the viscoelasticity. The stress relaxation experiments are used to extract weighted time scales via a generalized Maxwell viscoelastic model and those time scales directly relate to the molecular features. Experimental uniaxial tensile stress-strain data are recreated in a finite element analysis (FEA) using the Ogden hyperelastic model. The Maxwell and Ogden model parameters are input into an FEA to predict the strain recovery difference and then the curvature change with time where the predictions agree well with experiments. The polymer physics approach used to design shape-changing materials relies on the intrinsic behavior of the polymer molecules after applying and releasing a deformation. This is in contrast to, for instance, the diffusion of a small molecule solvent such as water into the system then back out again. In that system, the time scale for shape change is the solvent diffusion time, which depends on temperature, the interaction free energy between the solvent and polymer, and the sample thickness. The bilayers are made of proteins and sugars and are generally regarded as safe, edible, and biodegradable. Thus, stimuli-responsive shape changing materials can be created that are bioinspired and biobased.

4. **Materials and Methods**

*Materials*: The materials are inspired by gummy candy, which is hyperelastic and when deformed and released, will return to its original shape over time. Table 1 lists the recipes for a low elastic recovery (ER-low) and a high elastic recovery (ER-high) layer. Gelatin, sugar (sucrose), and light corn (glucose) syrup (all Kroger Supermarket brand), glycerol (BDH1172 from VWR) and citric acid (Roots Circle from Amazon) are all used as-received. Gelatin and de-ionized water are mixed, sealed in a plastic bag, and "bloomed" in a water bath at 60 °C for 30 minutes. The sugar and corn syrup or sugar, corn syrup, and glycerol are mixed and heated to 100 °C (with glycerol) or 110 °C (without glycerol) where the mixture begins to boil and clarify. Upon reaching a boil, the temperature is reduced to 60 °C and the bloomed gelatin, which is now a liquid, is added to the sugar mixture. Care is taken not to add bubbles from the gelatin into the mixture. The mixture is mixed slowly to not introduce any bubbles. Any formed bubbles are vacuumed out and the mixture remains liquid as long as it is maintained above 37 °C, the melting point of the gelatin. The mixture is finally poured into a 17.8 x 17.8 cm silicone mold. For tensile testing, 200 g of each formula is prepared resulting in samples about 4 mm thick. The sample is allowed to set at room temperature for 30 minutes followed by refrigeration for 24 hours. To prepare bilayers, 100 g of ER-low is first poured into the mold and allowed to set for 30 minutes at room temperature. Then, 110 g of ER-high is poured on top and allowed to set for another 30 min at room temperature. After the 2$^{nd}$ 30 minute setting, the bilayer is refrigerated for 24 hours.

**Table 1. Gelatin viscoelastic layer formulations.**

| Component | ER-low (mass fraction) | ER-high (mass fraction) |
|---|---|---|
| gelatin | 0.101 | 0.101 |
| water | 0.203 | 0.203 |
| sugar | 0.324 | 0.162 |
| corn syrup | 0.351 | 0.340 |
| glycerol | 0.000 | 0.173 |
| citric acid | 0.021 | 0.021 |

Samples are dried to a final mass, m, of 90% of the original mass, $m_0$, (normalized mass of $m/m_0$=0.90 is 49.3% water loss, the only component that is evaporated, resulting in a new water mass fraction of 0.114) and 85% of the original mass ($m/m_0$=0.85 is 73.9% water loss with a new water mass fraction of 0.062, Table S1). This is done in the laboratory under a fume hood and the same rate of water loss is achieved no matter the ambient relative humidity. It is easy to see the layers under a stereoscope and measure the thickness of each and the measured thickness is the same as that calculated from the mass, density, and surface area. The layers adhere very well and no delamination is ever observed.

*Bilayer Curvature*: Bilayers are cut into quasi-1D strips ($L_0 > W_0 \gg H_0$) of original length, $L_0 = 10$ cm, width, $W_0 = 1$ cm, and thickness, $H_0 = 0.45$ cm (after drying to $m/m_0 = 0.90$) or $H_0 = 0.42$ cm (after drying to $m/m_0 = 0.85$). The bilayers are pulled to a desired strain and released. The process is recorded next to a scale in two directions: from the side (parallel view) and bottom (perpendicular view, Figure S1). The videos are imported into Fiji image analysis software and the maximum curvature at the bilayer midplane found as a function of time, $\kappa(t)$, as previously reported.[4j, 4k, 10a]

*Stress-Strain and Stress Relaxation*: ASTM D412 Die D tensile specimens are used for strain to break, stress relaxation, and modified creep recovery experiments. For stress-strain experiments, samples are uniaxially deformed in tension at a crosshead speed of $r = 8.33$ mm s$^{-1}$ (500 mm min$^{-1}$) on a Texture Analyzer TA.HD*Plus* (Texture Technologies Corp.) equipped with a 5-kg load cell and self-tightening crosshatched grips. Engineering or nominal stress, $\sigma$, and strain, $\varepsilon$, are found from the original sample cross-sectional area and the crosshead displacement relative to the sample gage length, respectively. 7-8 samples are tested for each formulation. Stress relaxation experiments are conducted at the same crosshead speed by stretching the sample to a desired strain and then holding for up to 3 minutes, by which time the sample has reached a long period of constant stress (Figure S2c,d). 3-6 samples are tested at each applied strain. Results are reported as an average ± standard deviation.

*Modified Creep Recovery*: To mimic the act of pulling and releasing a bilayer to achieve curvature then allowing it to freely recover back to its original shape, a modified creep recovery experiment is developed that does not follow a steady-state creep experiment. Instead, the sample is stretched to a constant strain rather than stress and then released although approximately the same stress is achieved each time at the given strain. Modified creep recovery experiments are performed on a Texture Analyzer TA.XT*Plus* (Texture Technologies Corp.) also using a 5-kg load cell and the same grips. This instrument has a faster motor and PID controller than the TA.HD*Plus* but a smaller frame, making it more suitable for smaller strain elastomer testing for the given sample geometry. Only stress-strain tensile tests went to high strains and the bilayers are never tested above 100% strain. No differences are observed on samples tested on both instruments. Samples of the individual layers are strained at 8.33 mm s$^{-1}$ to the desired strain, held for 0.5 seconds, then the stress is released and the crosshead follows and records the sample retraction with time for 60 seconds. 3-6 samples are tested at each applied strain and results reported as an average ± standard deviation.

*Constitutive Model Parameter Identification*: To obtain the hyper-viscoelastic constitutive model parameters of each layer, an inverse parameter identification program is developed (Section S1, Figure S2). The stress relaxation data is fit by the stress history function derived from the integral constitutive equation of the generalized Maxwell model (Equation S4). Finite element analysis (FEA) modeling is employed to simulate the standard dog-bone specimen in the uniaxial tensile tests and the modified creep recovery tests using the FEAP v8.6 package.[21] Material hyperelastic behavior is modeled by the Ogden model (Equation S7) and the viscoelastic properties are considered through the three-term Prony's series in FEA (Equation S1). The objective of the program is to optimize parameter values of the Ogden model and Prony's series with the aim of minimizing errors in the strain recovery history between FEA modeling and modified creep recovery experiments, errors in the stress-strain relationships between FEA modeling and uniaxial tensile experiments, and errors in the solutions of Equation S4 when compared to stress relaxation data. The identified parameters in the Prony's series and in the Ogden model are given in Table S2 and S3, respectively.

*Bilayer Finite Element Analysis (FEA)*: Nonlinear finite element analyses are conducted to understand bilayer curvature. The FEA model is established by employing quadratic 27-node solid elements with a mixed finite element formulation (Figure S3). The hyper-viscoelastic characteristics of the two layers are represented using the Ogden model combined with a three-term Prony's series within the FEA framework. The bilayer is subjected to the same vertical displacement load used in the modified creep recovery test, with a fully clamped boundary condition at the top (Figure S2a, S3). A small horizontal displacement load ($D_v$=0.3 mm) is introduced during the stretching process to account for perturbation effects stemming from eccentric loading during stretching. Gravity is modeled as a body force with material density assumed to be 1.0 g cm$^{-3}$ and the gravitational acceleration of 9.81 m s$^{-2}$. To ensure convergence, an extremely small time step of $\Delta t=10^{-5}$ s is employed for modeling the bilayer immediately after the displacement load is removed. Once the shape recovery of the bilayer is established, converged solutions can be efficiently obtained using larger time steps. The simulations continue until the bilayer reaches a state of nearly zero curvature.

**Supporting Information**

Supporting Information is available from the Wiley Online Library or from the author.

**Acknowledgements**

We are grateful to Sam Garbera, Rachel Tschetter, and Joe Venturato for help with the curvature and stress-strain experiments.

**Conflict of Interest**

The authors declare no conflict of interest.

Received:

Revised:

Published online:References

[1] a)E. Reyssat, L. Mahadevan, *J. R. Soc. Interface* **2009**, 6, 951; b)K. Song, E. Yeom, S.-J. Seo, K. Kim, H. Kim, J.-H. Lim, S. Joon Lee, *Sci. Rep.* **2015**, 5, 9963; c)A. Le Duigou, M. Castro, *Sci. Rep.* **2016**, 6, 18105.

[2] a)S. Armon, H. Aharoni, M. Moshe, E. Sharon, *Soft Matter* **2014**, 10, 2733; b)S. Armon, E. Efrati, R. Kupferman, E. Sharon, *Science* **2011**, 333, 1726; c)D. Evangelista, S. Hotton, J. Dumais, *J. Exp. Biol.* **2011**, 214, 521; d)R. Elbaum, Y. Abraham, *Plant Sci.* **2014**, 223, 124.

[3] a)S. J. Gerbode, J. R. Puzey, A. G. McCormick, L. Mahadevan, *Science* **2012**, 337, 1087; b)M. H. Godinho, J. P. Canejo, G. Feio, E. M. Terentjev, *Soft Matter* **2010**, 6, 5965; c)A. Goriely, M. Tabor, *Phys. Rev. Lett.* **1998**, 80, 1564.

[4] a)Z. Hu, X. Zhang, Y. Li, *Science* **1995**, 269, 525; b)P. D. Topham, J. R. Howse, C. J. Crook, S. P. Armes, R. A. L. Jones, A. J. Ryan, *Macromolecules* **2007**, 40, 4393; c)L. Ionov, *Soft Matter* **2011**, 7, 6786; d)S.-W. Lee, J. H. Prosser, P. K. Purohit, D. Lee, *ACS Macro Lett.* **2013**, 2, 960; e)Y. Gu, N. S. Zacharia, *Adv. Funct. Mater.* **2015**, 25, 3785; f)X. Li, X. Cai, Y. Gao, M. J. Serpe, *J. Mater. Chem. B* **2017**, 5, 2804; g)A. H. Velders, J. A. Dijksman, V. Saggiomo, *Appl. Mater. Today* **2017**, 9, 271; h)Y. Cheng, K. Ren, D. Yang, J. Wei, *Sensors Actuators B: Chem.* **2018**, 255, 3117; i)A. H. Nourian, A. Amiri, N. Moini, M. Baghani, *Smart Mater. Struct.* **2020**, 29, 105001; j)L. E. Hanzly, N. Chauhan, J. R. Barone, *Smart Mater. Struct.* **2022**, 31, 085005; k)L. E. Hanzly, K. A. Kristofferson, N. Chauhan, J. R. Barone, *Green Mater.* **2021**, 9, 157; l)D. Morales, E. Palleau, M. D. Dickey, O. D. Velev, *Soft Matter* **2014**, 10, 1337.

[5] a)N. Bassik, B. T. Abebe, K. E. Laflin, D. H. Gracias, *Polymer* **2010**, 51, 6093; b)G. Stoychev, N. Puretskiy, L. Ionov, *Soft Matter* **2011**, 7, 3277; c)G. Stoychev, S. Zakharchenko, S. Turcaud, J. W. C. Dunlop, L. Ionov, *ACS Nano* **2012**, 6, 3925; d)T. S. Shim, S.-H. Kim, C.-J. Heo, H. C. Jeon, S.-M. Yang, *Angew. Chem. Int. Ed.* **2012**, 51, 1420; e)J. Kim, J. A. Hanna, M. Byun, C. D. Santangelo, R. C. Hayward, *Science* **2012**, 335, 1201; f)J. Kim, J. A. Hanna, R. C. Hayward, C. D. Santangelo, *Soft Matter* **2012**, 8, 2375; g)G. Qi, K. D. Conner, H. J. Qi, L. D. Martin, *Smart Mater. Struct.* **2014**, 23, 094007; h)A. Sydney Gladman, E. A. Matsumoto, R. G. Nuzzo, L. Mahadevan, J. A. Lewis, *Nature Mater.* **2016**, 15, 413; i)M. Tyagi, G. M. Spinks, E. W. H. Jager, *SmartWe are grateful to Sam Garbera, Rachel Tschetter, and Joe Venturato for help with the curvature and stress-strain experiments.

**Conflict of Interest**

The authors declare no conflict of interest.

Received:

Revised:

Published online:

**Figures**

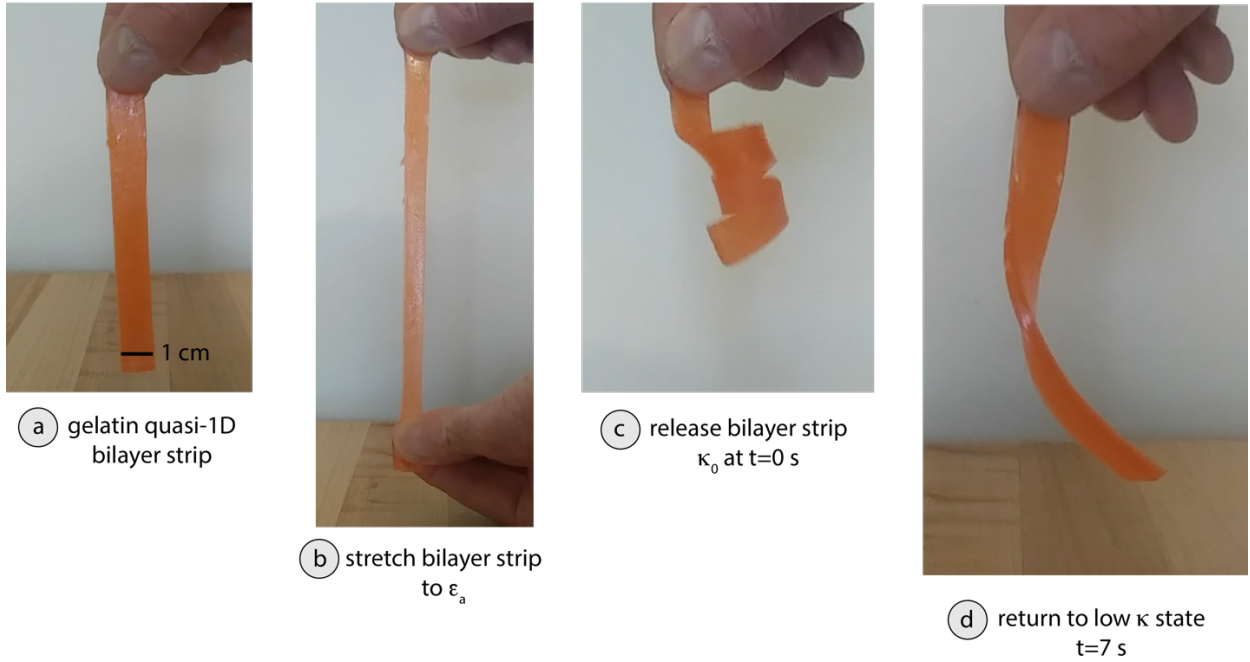

**Figure 1.** Stretch and release sequence for (a) quasi-1D bilayers of $L_0$=10 cm, $W_0$=1 cm, and $H_0$=0.45 cm and 0.114 mass fraction water. (b) Bilayer is stretched to $\varepsilon_a$=1.00 cm/cm and (c) released where it twists into a helix of zero time or maximum curvature, $\kappa_0$. (d) The bilayer eventually returns to a low $\kappa$ state where the process can be repeated.

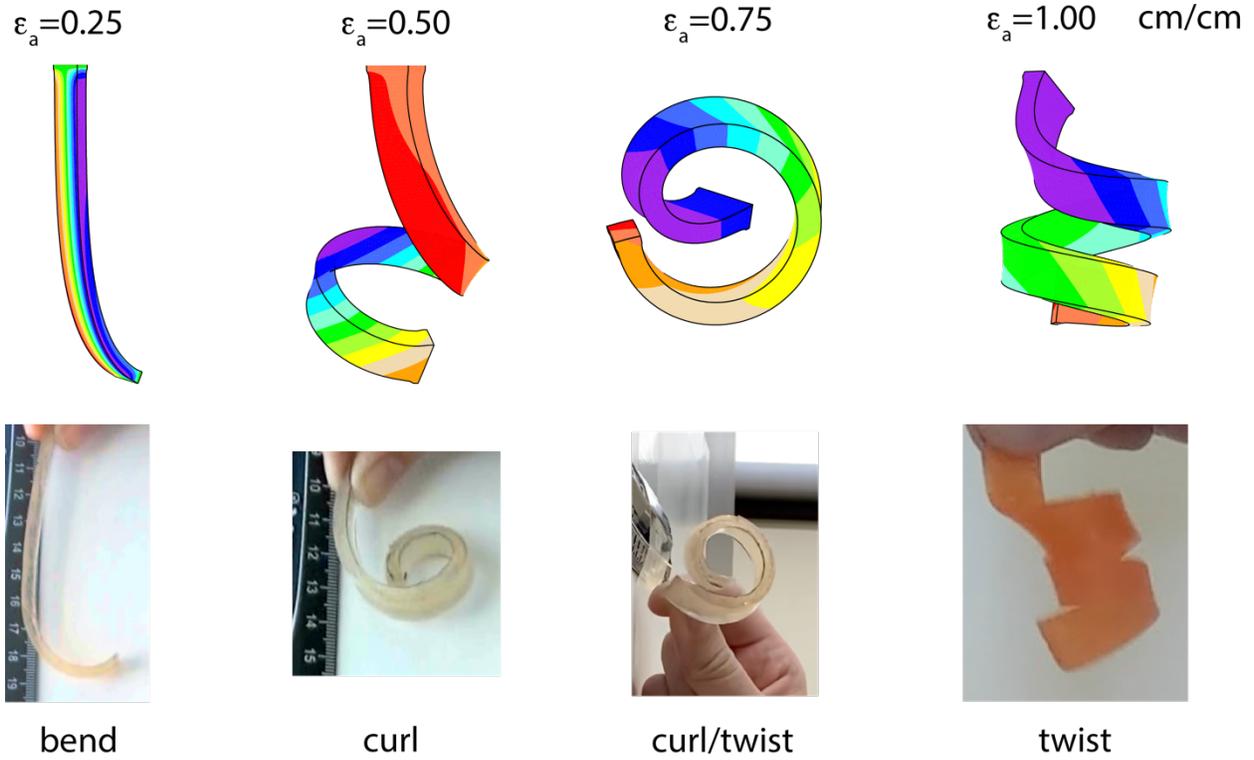

**Figure 2.** Examples of initial curvature showing bending, curling, and twisting with increasing applied strain. Also shown are the FEA predicted shapes.

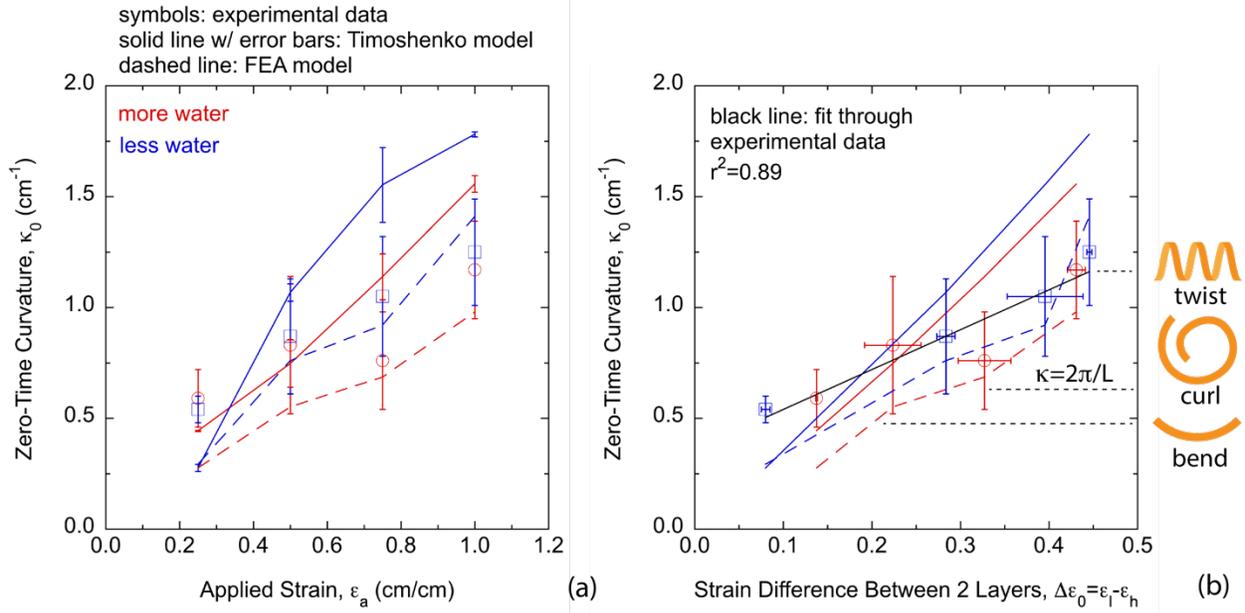

**Figure 3.** The maximum curvature, $\kappa_0$, is reached immediately upon release at time, t=0 s. $\kappa_0$ plotted vs. (a) applied strain, $\varepsilon_a$, and (b) the initial strain difference between the 2 layers in the bilayer, $\Delta\varepsilon_0=\varepsilon_l-\varepsilon_h$ (defined in Figure 3). The error bars for the Timoshenko model are left out of (b) for clarity. "More" and "less" water are 0.114 and 0.062 mass fraction water, respectively.

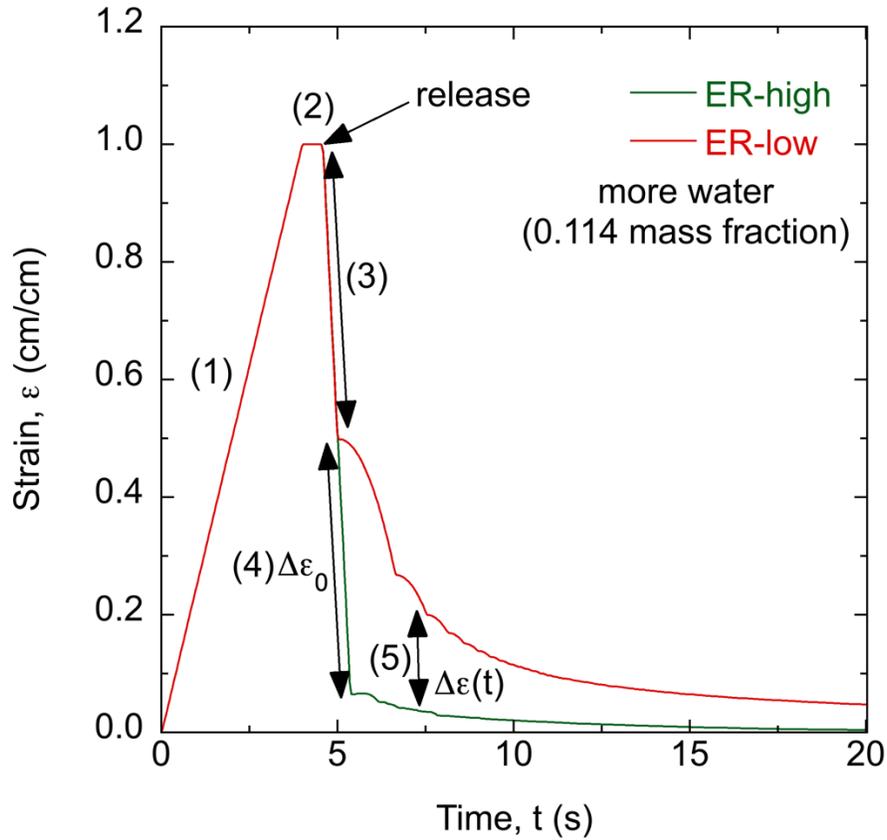

**Figure 4.** Example of the modified creep recovery experiment showing: (1) loading at 8.33 mm/s to a strain of 1.0 cm/cm, (2) holding for 0.5 s then releasing, (3) immediate equal recovery (elastic recoil) of each layer until (4) where the ER-low layer begins strain-dependent recovery but the ER-high layer is still immediately recovering strain. $\Delta\varepsilon_0$ is the initial strain difference yielding the initial maximum curvature, $\kappa_0$. (5) Soon after, the ER-high layer begins to recover strain with time. Both layers recover strain differently over time, $\Delta\varepsilon(t)$, resulting in the curvature decrease with time.

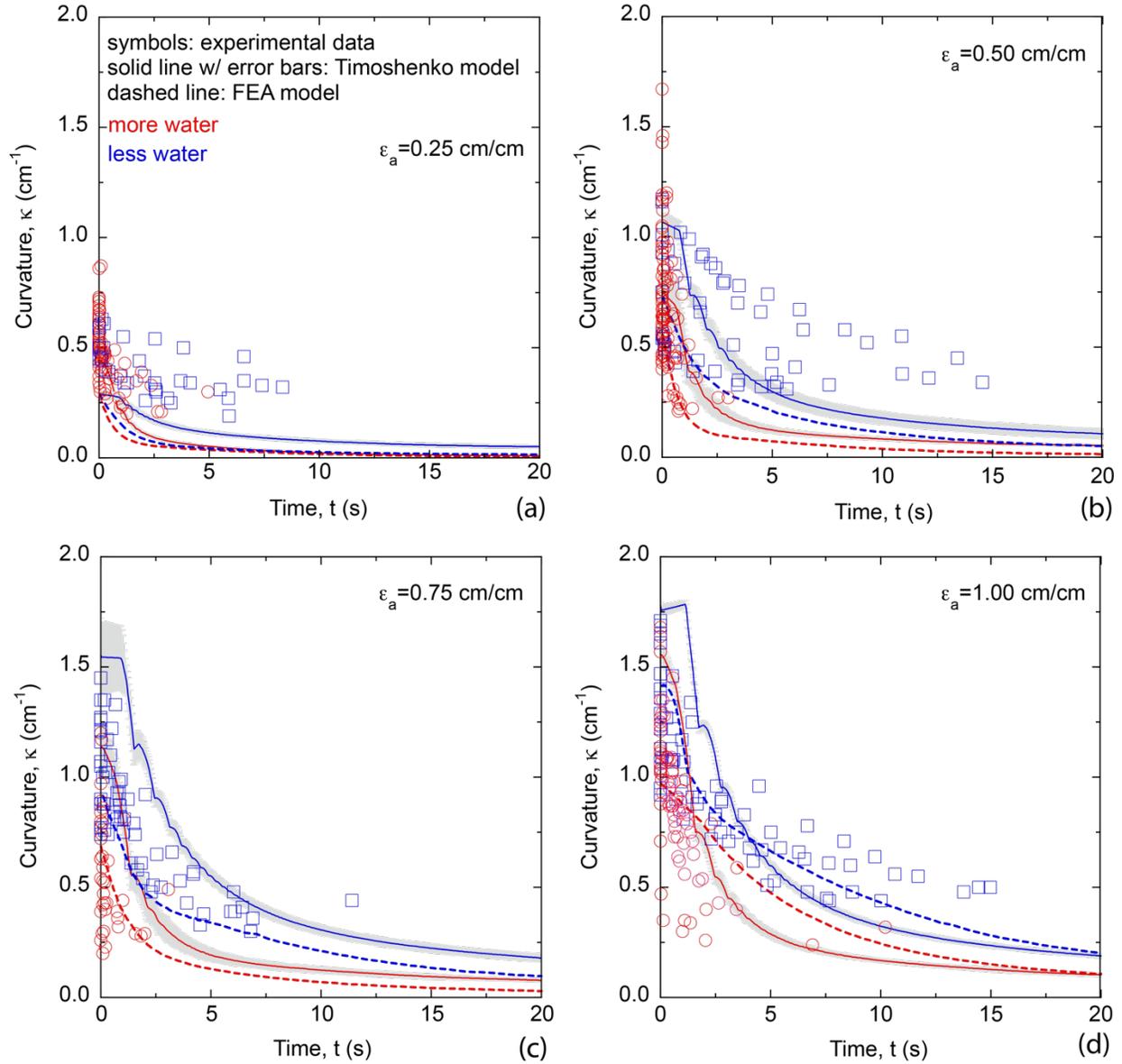

**Figure 5.** Curvature, $\kappa$, changes as a function of time for applied strains, $\varepsilon_a$, of (a) 0.25 cm/cm, (b) 0.50 cm/cm, (c) 0.75 cm/cm, and (d) 1.00 cm/cm. "More water" and "less water" bilayers contain 0.114 and 0.062 mass fraction water, respectively. The solid lines with gray error bars are the Timoshenko model results based on the experimentally determined strain recovery difference between the two layers, $\Delta\varepsilon(t)$. The dashed lines are the FEA model results based on fitting the stress-strain, $\sigma$-$\varepsilon$, experimental data to the Ogden hyperelastic model and the stress relaxation, $\sigma(t)$, experimental data to a 3-term Prony series.

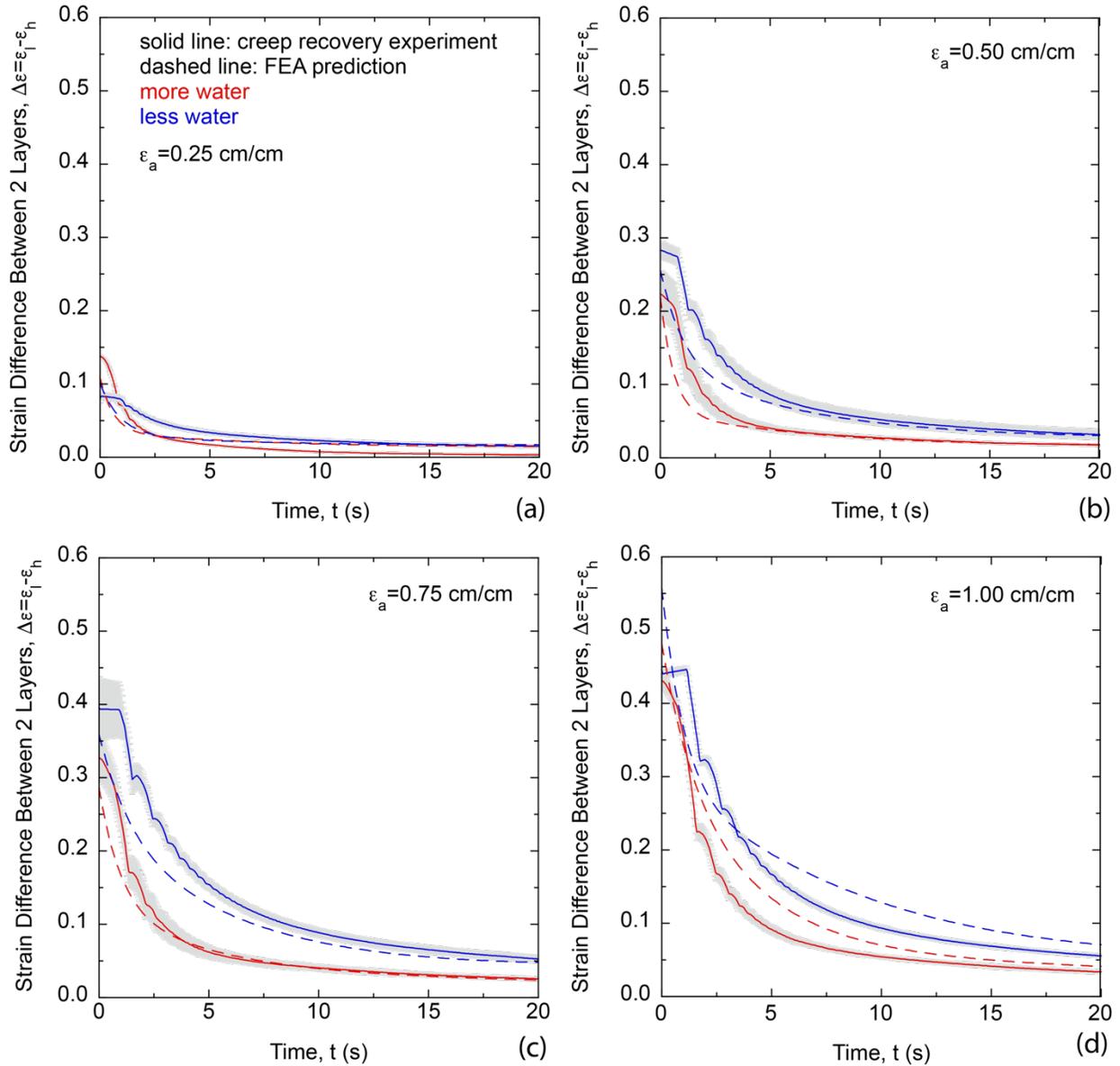

**Figure 6.** Strain difference between the 2 layers in the bilayer, $\varepsilon$, changes as a function of time for applied strains, $\varepsilon_a$, of (a) 0.25 cm/cm, (b) 0.50 cm/cm, (c) 0.75 cm/cm, and (d) 1.00 cm/cm. "More water" and "less water" bilayers contain 0.114 and 0.062 mass fraction water, respectively. The solid lines are the results from the modified creep experiment and the dashed lines are the predicted strain difference from FEA.

Wenya Shu, C. Nadir Kaplan*, Justin R. Barone*

**Dynamic, viscoelasticity-driven shape change of elastomer bilayers**

Shape-morphing materials can have tailored temporal responses based on the inherent viscoelasticity of the constituent polymers. Upon stretching and releasing, gelatin elastomer bilayers with layers of varying viscoelasticity can bend, curl, and twist into 3D shapes that will relax back to the original flat shape on a time scale determined by the polymer molecular features.

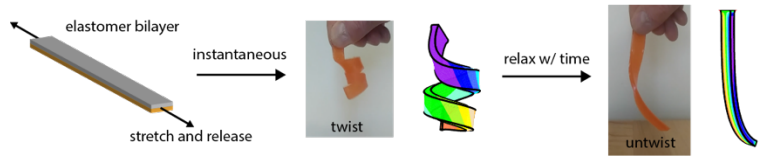